\documentclass[twocolumn,prl,aps,showpacs,superscriptaddress,floatfix]{revtex4}

\usepackage{graphicx}

\begin{document}


\title{Evidence for Two Different Solid Phases of Two Dimensional Electrons in High Magnetic Fields}

\author{Yong P.~Chen}
\affiliation{Department of Electrical Engineering, Princeton University,
Princeton, NJ 08544}
\affiliation{National High Magnetic Field Laboratory, 1800 E.~Paul Dirac Drive, 
Tallahassee, FL 32310}
\author{R.~M.~Lewis}
\affiliation{National High Magnetic Field Laboratory, 1800 E.~Paul Dirac Drive, 
Tallahassee, FL 32310}
\affiliation{Department of Electrical Engineering, Princeton University,
Princeton, NJ 08544}
\author{L.\ W.\
Engel}
\affiliation{National High Magnetic Field Laboratory, 1800 E.~Paul Dirac Drive, 
Tallahassee, FL 32310}
\author{D.\ C.\ Tsui}
\affiliation{Department of Electrical Engineering, Princeton University,
Princeton, NJ 08544}
\author{P.\ D.\ Ye}
\altaffiliation[Current address: ]{Agere Systems, 555 Union Blvd., Allentown PA 18109}
\affiliation{National High Magnetic Field Laboratory, 1800 E.~Paul Dirac Drive, 
Tallahassee, FL 32310}
\affiliation{Department of Electrical Engineering, Princeton University,
Princeton, NJ 08544}
\author{Z.~H.~Wang}
\affiliation{Department of Physics, Princeton University,
Princeton, NJ 08544}
\affiliation{National High Magnetic Field Laboratory, 1800 E.~Paul Dirac Drive, 
Tallahassee, FL 32310}
\author{L.\ N.\ Pfeiffer}
\affiliation{Bell Laboratories, Lucent Technologies, Murray Hill, NJ 07974}
\author{K.\ W.\ West}
\affiliation{Bell Laboratories, Lucent Technologies, Murray Hill, NJ 07974}

\date{\today}

\begin{abstract}
We have performed RF spectroscopy on very high quality two dimensional electron systems in the high magnetic field insulating phase, usually associated with a Wigner solid (WS) pinned by disorder. We have found two different resonances in the frequency dependent real diagonal conductivity spectrum and we interpret them as coming from \textit{two} different pinned solid phases (labeled as ``WS-A" and ``WS-B"). The resonance of WS-A is observable for Landau level filling $\nu$$<$2/9 (but absent around the $\nu$=1/5 fractional quantum Hall effect (FQHE)); it then \textit{crosses over} for $\nu$$<$0.18 to the different WS-B resonance which dominates the spectrum at $\nu$$<$0.125.  Moreover, WS-A resonance is found to show dispersion with respect to the size of transmission line, indicating that WS-A has a large correlation length (exceeding $\sim$100 $\mu$m); in contrast no such behavior is found for WS-B.  We suggest that quantum correlations such as those responsible for FQHE may play an important role in giving rise to such different solids.

\end{abstract}

\maketitle

In 1934, Wigner \cite{wigner} proposed that electrons can crystallize into a solid when their (Coulomb) interaction energy dominates over their kinetic energy. In two dimensions, it is expected \cite{lozovik} that formation of such Wigner solid (WS) can be facilitated by a sufficiently strong perpendicular magnetic field ($B$).  On the other hand, a two dimensional electron system (2DES) with areal density $n$ can condense into quantum Hall (QHE) states \cite{prange} with dissipation-free transport at a series of integer or fractional Landau filling factors $\nu$=$(h/e)(n/B)$, where $h/e$ is the Dirac flux quantum. Calculations \cite{wctrans} predicted the transition from fractional QHE series (which are incompressible quantum liquids) to WS to occur around $\nu$=1/5.  DC transport studies \cite{willett,jiang} on high quality (low disorder) samples at the lowest temperatures have found $\nu$=1/5 as the lowest $\nu$ fractional QHE state (FQHE), before the 2DES enters an insulating phase at higher $B$. Early experiments \cite{wcrev} on this high $B$ insulating phase (HBIP) were interpreted as consistent with an electron solid pinned by disorder. Here, we report radio frequency (RF) spectroscopy experiments on extremely low disorder 2DES and present evidence that such HBIP may consist of not just one, but \textit{two} different solid phases.

Using RF and microwave spectroscopy, recent experiments \cite{peide} have measured a high quality 2DES down to $\nu$ as small as $\sim$1/25, and observed a single sharp resonance in the frequency ($f$) dependent real diagonal conductivity (Re[$\sigma_{xx}(f)$]) of 2DES in HBIP. Such a resonance (not observed, for example, in FQHE liquids) has been taken as a signature of solid and interpreted as due to the ``pinning mode" (the disorder gapped lower branch of the magnetophonon) \cite{fulee,normand,fertig,fogler,chitra} of WS crystalline domains oscillating collectively within the disorder potential. 

 In our experiments, we have measured even lower disorder 2DES and observed \textit{two} different resonances in different regimes of HBIP, with one resonance crossing over to the other as $\nu$ is reduced (by increasing $B$). We interpret the two resonances as coming from \textit{two} different solid phases pinned by disorder. 

	The 2DES samples we have used are fabricated from two very high quality GaAs/AlGaAs/GaAs quantum well (QW) wafers grown by molecular beam epitaxy.   Data from three samples will be presented. Sample 1 contains a 50nm-wide QW with $n$=1.0$\times$10$^{11}$cm$^{-2}$ and mobility $\mu$$\sim$1$\times$10$^7$cm$^2$/Vs. Sample 2a and 2b are from the other wafer, each containing a 65nm-wide QW with $n$=5.1$\times$10$^{10}$cm$^{-2}$ and 
$\mu$$\sim$8$\times$10$^6$cm$^2$/Vs.  

\begin{figure}[tb!]
\includegraphics[width=8cm]{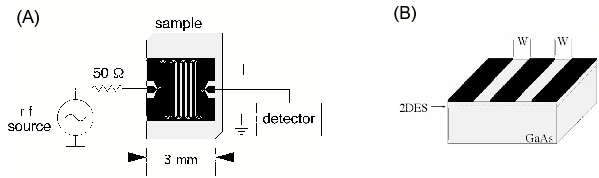}
\caption{\label{fig:fig1} {\bf (A)} Scheme of microwave circuit typically used in our experiment. Sample size is $\sim$3$\times$5 mm.  Dark regions on sample surface represent metal films deposited to make the CPW.  The dimensions of the CPW can vary, but have been carefully designed to match the 50$\Omega$ characteristic impedance.  \textbf{(B)} Magnified (not to scale) local cross section of sample with CPW, where $w$ is the width of each slot region.  The 2DES resides in a symmetric AlGaAs/GaAs/AlGaAs QW and is 0.4-0.5 $\mu$m under the surface.  Sample substrate is GaAs.  }   
\end{figure}

We have deposited on the surface of each sample a metal film coplanar waveguide (CPW) similar to the ones used in previous experiments measuring microwave conductivity of 2DES \cite{peide,engel,bubble,iqhewc}. A typical measurement circuit is shown schematically in Fig. ~\ref{fig:fig1}(A) and a magnified (not to scale) local cross section of the sample near the CPW is shown in Fig. ~\ref{fig:fig1}(B).  A network analyzer generates and detects the RF signal, which propagates along the CPW and couples capacitively to the 2DES.  The CPW confines the electric field ($E$) mainly in each slot region of width $w$ (shown in Fig. ~\ref{fig:fig1}(B)), giving $E$ a step function profile (neglecting edge effects related to the 2DES \cite{fogler}), therefore introducing a finite wavevector through the dominant Fourier component $q$$\sim$$\pi/w$. The relative power absorption ($P$) by the 2DES is measured.  Under conditions of sufficiently high $f$ and low 2DES conductivity, no reflections at ends of CPW, and when 2DES is in its long wave length limit \cite{engel}, 
$P=\exp((2lZ_0/w)\mathrm{Re}[\sigma_{xx}])$, where $l$ is the total length of the CPW, $Z_0$ its characteristic impedance (50$\Omega$), and Re[$\sigma_{xx}$] is the real part of diagonal conductivity of 2DES. We cast our measured $P$, even though the 2DES in our experiments is not in its long wavelength limit as will be seen, into a real diagonal conductivity which we define Re[$\sigma ^{c}_{xx}$]=$(w/2lZ_0)\ln(P)$. We used the meandering CPWÕs \cite{strcpw} to obtain larger geometric factors ($2l/w$) therefore increasing the strength of absorption signal ($P$). Samples are immersed in $^3$He-$^4$He mixture in a dilution refrigerator and placed in a perpendicular $B$. 
Measurements are done in the low RF power limit, by reducing power till absorption $P$ no longer changes.

\begin{figure}[tb!]
\includegraphics[width=8.5cm]{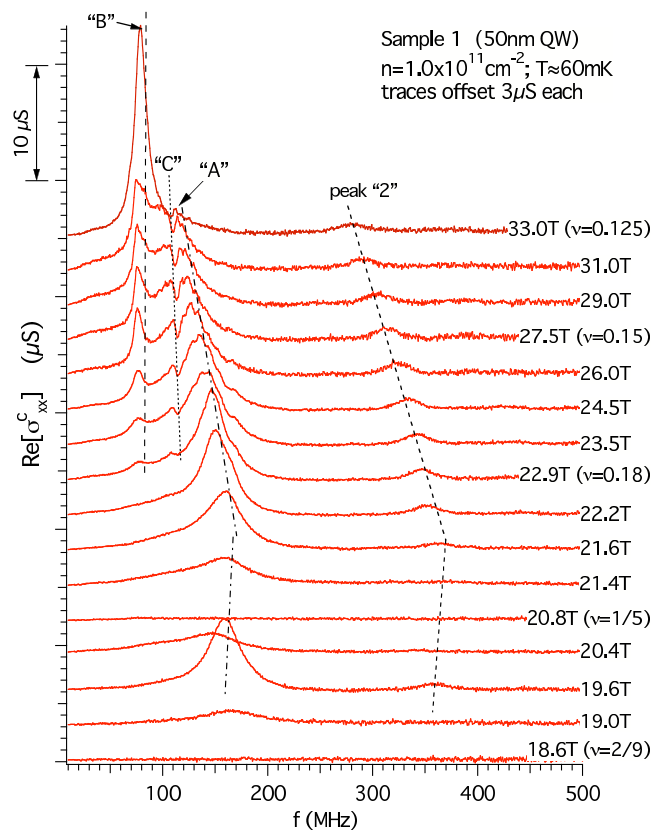}
\caption{\label{fig:fig2}Sample 1: Re[$\sigma ^{c}_{xx}(f)$]  spectra at various $B$, in increasing order from $B$=18.6T (bottom) to 33T (top). Adjacent traces are offset for 3$\mu$S from each other for clarity.  Magnetic fields (and selected $\nu$'s) are labeled at right.  Measurements were performed at $T$ $\sim$60 mK. From left to right, the long dashed, dotted, dot-dashed, and short dashed lines are guides to the eye, corresponding to peaks ``B", ``C", ``A" and ``2" respectively, as explained in the text. 
}
\end{figure}

Fig. ~\ref{fig:fig2} shows Re[$\sigma ^{c}_{xx}(f)$] spectra measured at various $B$ from sample 1, at the temperature ($T$) $\sim$60 mK. The traces are displayed in increasing order of $B$ from bottom (taken at 18.6T) to top (at 33T) and offset 3$\mu$S from each other for clarity. The spectrum is flat at $B$=18.6T, corresponding to the $\nu$=2/9 FQHE liquid state.  Upon increasing $B$, a clear resonance (with peak frequency ($f_\mathrm{pk}$) $\sim$150 MHz) can be observed; the resonance is interrupted briefly (with flat spectrum) near $\nu$=1/5 FQHE liquid then reappears at higher $B$. This resonance, reentrant \cite{rip} around $\nu$=1/5, will be referred to as peak ``A" hereafter.  At 22.9T, $\nu$$\sim$0.18, another resonance, labeled as ``B", starts to appear ($f_\mathrm{pk}$ $\sim$80 MHz).  Further increasing magnetic field, resonance ``B" grows while ``A" continues to evolve but eventually weakens. By 33T ($\nu$=0.125), resonance ``B" dominates the spectrum and ``A" nearly disappears.  From 22.9T to 33T ($\nu$ from 0.18 to 0.125) the spectra display a clear crossover from ``A" to ``B"; moreover, in this crossover region they show complicated structures, for example an intermediate peak like the one labeled ``C" appearing between ``A" and ``B".  

We have also observed higher lying but relatively weak resonances such as the one labeled as peak ``2" in the figure.  They show qualitative similarities with ``A" (for example, the dependence on magnetic field) but do not appear to fit simple harmonics of ``A". Details of them will be discussed in a future publication. 

\begin{figure}[tb!]
\includegraphics[width=8.5cm]{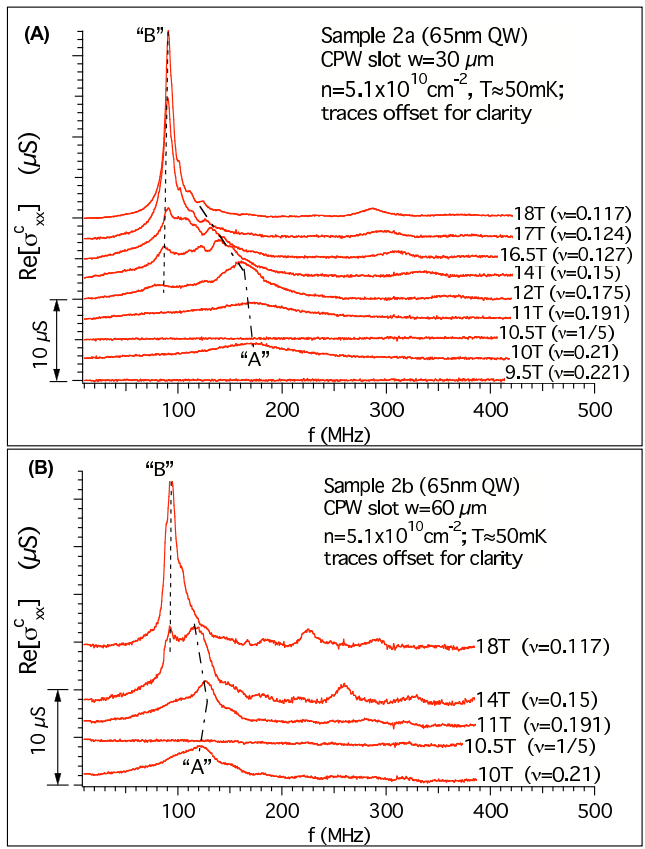}
\caption{\label{fig:fig3}{\bf (A)} Sample 2a: Re[$\sigma ^{c}_{xx}(f)$] spectra at various $B$, in increasing order from 9.5T (bottom) to 18T (top). Adjacent traces are appropriately offset for clarity.  Values of $B$ and  $\nu$ for each trace are labeled at right.  Measurements were performed at $T$$\sim$50 mK. Dash and dot-dashed lines are guides to the eye and correspond to resonances ``A" and ``B", similarly defined as in Fig.~\ref{fig:fig2}.  Compared to sample 1, sample 2a has about half the density, and the crossover from resonance ``A" to ``B" also occurs at about half the $B$, resulting in the similar $\nu$ range of crossover.  \textbf{(B)} Spectra at five representative magnetic fields measured from sample 2b. Sample 2b, which has a CPW with $w$=60 $\mu$m, is from the same wafer as sample 2a (which has a CPW with $w$=30 $\mu$m) and measured at $\sim$50 mK in a separate cooldown which gave the same density.  Traces are appropriately offset for clarity. Compared to corresponding traces in \ref{fig:fig3}(A), peaks labeled as resonance ``B" are seen to occur at the same frequencies but those labeled as ``A" shift to lower frequencies.  The flat spectrum at $\nu$=1/5 ($B$=10.5T) is also shown. }
\end{figure}

	Fig. ~\ref{fig:fig3}(A) shows Re[$\sigma ^{c}_{xx}(f)$] spectra measured from sample 2a, in which we observe behavior similar to that of sample 1, with one resonance (``A") reentrant around $\nu$=1/5 crossing over to a different resonance (``B") dominating at sufficiently small $\nu$. We emphasize that, compared to sample 1, the crossover here occurs at much lower $B$, but in similar $\nu$ (from $\sim$0.18 to $\sim$0.125) range. The same $\nu$ range of crossover has also been found in another cooldown which gave a different density for sample 2a.

		A striking difference between resonances ``A" and ``B" is seen by comparing Fig.~\ref{fig:fig3}(A) to \ref{fig:fig3}(B), which shows the spectra measured at five representative magnetic fields using sample 2b. Sample 2b was cut from the same wafer with sample 2a and only differs in the slot width ($w$) of the CPW. Both samples show similar resonances ``A" and ``B", with similar $\nu$ range of crossover. However, going from $w$=30 $\mu$m (sample 2a) to 60 $\mu$m (sample 2b), we notice that $f_\mathrm{pk}$ of resonance ``A" shifts to lower value while $f_\mathrm{pk}$ of resonance ``B" is not affected; this is true even when resonances ``A" and ``B" coexist (for example, in the spectrum at 14T). Since $w$ introduces a finite wavevector in the measurement, we are apparently sensing the dispersion ($f_A(q)$) of resonance ``A" using samples with varying $w$. 

Our data thus reveal two distinct regimes in the HPIB characterized by two different resonances (``A" and ``B"): one at 2/9$<$$\nu$$<$0.18 (except for a narrow range around $\nu$=1/5) where only resonance ``A" has been observed; and another at $\nu$$<$0.125 (down to the smallest $\nu$ we have accessed) where the rather different resonance ``B" dominates \cite{blimit}. We interpret the two regimes as corresponding to two different (pinned) solid phases, hereafter referred to as ``WS-A" and ``WS-B" respectively, each being the preferred ground state in the respective $\nu$ range.  Because of interaction with disorder, either solid is pinned (thus insulating), and can support a similar pinning mode \cite{fulee,normand,fertig,fogler,chitra} that gives rise to the observed resonance. The $f_\mathrm{pk}$ of our resonances are nearly an order of magnitude lower than what previous experiments \cite{peide} observed \cite{cmppd}, probably due to significantly reduced pinning disorder in our samples. 

The striking crossover behavior, which we do not observe for $T$ above $\sim$130 mK \cite{tnote}, is consistent with a magnetic field induced phase transition from WS-A to WS-B and with coexistence of the two phases (at low $T$) in the transition regime (0.18$>$$\nu$$>$0.125), suggesting such transition would be first order.  The intermediate peak ``C" disappears at $\sim$100 mK, leaving only peaks ``A" and ``B" present in the spectra. Though peak ``C", like ``A" and ``B", is reproducible in different cooldowns of the same sample; we have sometimes noticed other delicate features that appear to depend on the way the sample is cooled (for example, peak ``B" sometimes briefly splits near $\nu$=0.125 before dominating the spectra at lower $\nu$'s). Such complicated behavior may reflect some delicate competition between multiple or intermediate phases in the transition regime. 

The apparent crossover from WS-A to WS-B is mainly controlled by Landau level filling $\nu$=$nh/eB$=2$(l_B/r)^2$, where the magnetic length $l_B=\sqrt{\hbar/eB}$ (which measures the size of electron wavefunction) and the mean separation between electrons $r=1/\sqrt{\pi n}$. This rules out the crossover being caused by interplay of  $l_B$ with, for example, some disorder length scale \cite{chitra} or as some $n$-induced transition, but rather points out the important role played by many-electron quantum correlations, dependent on $l_B/r$.

 	The dispersion behavior of resonance ``A" as seen in Fig.~\ref{fig:fig3} requires WS-A must have a correlation length larger than $w$ of the CPW; otherwise the pinned solid is effectively subjected to a uniform electric field, therefore can not couple to the finite $q$ introduced by $w$. Preliminary measurements on another sample with $w$=80 $\mu$m show that resonance ``A" continues to shift to lower $f$, implying a correlation length in WS-A at least on the order of $\sim$100 $\mu$m, which is two orders of magnitude larger than what the simple estimate used in \cite{peide}
 for a classical WS would give. 
 
	It has been thought that correlations responsible for FQHE can still be relevant \cite{wpan,yi,narev,csj} even in HBIP.  More specifically, theories \cite{yi,narev,csj} have considered different types of ``correlated" WS ($^m$CWS) made of  ``composite fermions" or  ``composite bosons", the quasiparticles (electrons bound with even or odd number ($m$) of flux quanta respectively) proposed to largely encapsulate the FQHE correlations \cite{jain}.  In this notation, $^\mathit{0}$CWS would be a WS made of ``bare" electrons, corresponding to the original case proposed in \cite{lozovik}. The theories \cite{yi,csj} have predicted a series of first order phase transitions among these different types of CWS as preferred ground states in different regimes of HBIP, thus offering an attractive interpretation for the different phases we observed as these different solids. So far, different theories \cite{yi,narev,csj} have favored, for example near $\nu$=1/5 (corresponding to our WS-A), different types of CWS's; and a detailed calculation of the dynamical responses of different CWS's pinned by disorder is not yet available to allow for a direct comparison with our observed resonances.  Although caution should be taken in comparing these theories (for ideal 2DES) to experiments on realistic samples, some predictions seem to be consistent with the experiments. For example, $\nu$=0.125, predicted to be a critical filling separating two different CWS phases \cite{yi,csj}, is in remarkable agreement with our phenomenological value below which WS-B resonance dominates.  Some other predictions, however, seem to be at odds with our observations. For example, $^\mathit{4}$CWS (favored in \cite{narev}) is predicted to significantly soften as $\nu$ approaches 1/5; for a weakly pinned solid \cite{fogler,chitra} this would result in increasing $f_\mathrm{pk}$ of the resonance, which is not the case we have observed. Generally, it is expected that even modest disorder may have significant influences on the various CWS phases \cite{narev}, for example, it may stabilize one CWS against another, consistent with the fact that previous microwave experiments \cite{peide} on a heterojunction sample with lower mobility ($\sim$5$\times$10$^6$cm$^2$/Vs) than our QW samples have observed only one resonance in HBIP. 

The spectroscopy measurements were performed at the National High Magnetic Field Laboratory, which is supported by NSF Cooperative Agreement No.~DMR-0084173 and by the State of Florida.  Financial support of this work was provided by AFOSR, DOE and the NHMFL-in house research program.  We thank Glover Jones, Tim Murphy and Eric Palm at NHMFL for experimental assistance.  We also thank                     G. Gervais, W. Pan, P. Stiles, and Kun Yang for inspirational discussions.


\begin{thebibliography}{}
\bibitem{strcpw}We have checked the consistency of our experimental findings using also straight line CPWs.
\bibitem{rip}Details of this resonance in the so called ``reentrant insulating phase" (RIP \cite{jiang}, 
for 1/5$<$$\nu$$<$2/9) and near the transition to $\nu$=1/5 and 2/9 FQHE liquids will be presented elsewhere.
\bibitem{blimit}Preliminary data from sample 2a with a relatively low as-cooled density  (4.6$\times$10$^{10}$cm$^{-2}$) and up to 33T shows resonance ``B" to continuously evolve down to  $\nu$$\sim$0.06.  No other resonances are observed in the spectrum at such small $\nu$.
\bibitem{cmppd}Previous experiments \cite{peide} found $f_\mathrm{pk}$$\sim$1 GHz or higher. Our samples show nearly flat spectrum at such high $f$.
\end{thebibliography}
\end{document}